\documentclass[authoryear,12pt,letter,3p]{jowarticle}
\usepackage{graphicx}
\usepackage{amsmath}
\usepackage{amssymb}
\usepackage{setspace}
\makeatletter
\renewcommand\section{\@startsection{section}{1}{\z@}{-3.25ex plus -1ex minus -.2ex}{1.5ex plus .2ex}{\normalsize\bf}}
\renewcommand\subsection{\@startsection{subsection}{2}{\z@}{-3.25ex plus -1ex minus -.2ex}{1.5ex plus .2ex}{\normalsize\bf}}
\renewcommand\subsubsection{\@startsection{subsubsection}{3}{\z@}{-3.25ex plus -1ex minus -.2ex}{1.5ex plus .2ex}{\normalsize\bf}}
\makeatother

\newtheorem{thm}{Theorem}[section]
\newtheorem{cor}[thm]{Corollary}

\newtheorem{defn}[thm]{Definition}

\numberwithin{equation}{section}

\newcommand{\ket}[1]{|#1\rangle}
\newcommand{\bra}[1]{\langle #1|}

\begin{document}
\begin{frontmatter}
\title{The Scope and Generality of Bell's Theorem}
\author{James Owen Weatherall}\ead{weatherj@uci.edu}
\address{Department of Logic and Philosophy of Science\\ University of California, Irvine, CA 92697}
\begin{abstract}I present what might seem to be a local, deterministic model of the EPR-Bohm experiment, inspired by recent work by Joy Christian, that appears at first blush to be in tension with Bell-type theorems.  I argue that the model ultimately fails to do what a hidden variable theory needs to do, but that it is interesting nonetheless because the way it fails helps clarify the scope and generality of Bell-type theorems.  I formulate and prove a minor proposition that makes explicit how Bell-type theorems rule out models of the sort I describe here.\end{abstract}
\end{frontmatter}

\doublespacing

\section{Introduction}

Bell-type theorems are usually understood to rule out a class of hidden variable interpretations of quantum mechanics that are, in a certain precise sense, local and deterministic.  These theorems come in a variety of forms, with different characterizations of what local and deterministic are meant to amount to.  In the simplest case, Bell-type theorems show that any hidden variable model satisfying some set of conditions fails to agree with the quantum mechanical predictions for the outcomes of a simple experiment, often called the EPR-Bohm experiment.  When one performs the suggested experiments, meanwhile, the results agree with quantum mechanics.  The upshot appears to be that no local, deterministic hidden variable model can correctly reproduce the measurement outcomes of a certain class of experiments.\footnote{Bell himself took the upshot to be that no \emph{local} model---deterministic or otherwise---could reproduce the measurement outcomes of such experiments \citep[see][]{Bell1965,Bell1966}.  For an excellent overview of the state of the art on Bell-type theorems, see \citet{Shimony}.  See also \citet{Jarrett1, Jarrett2} and \citet{MalamentBell} for particularly clear expositions of how Bell-type theorems work, with the relevant assumptions stated as precisely as possible.  An insightful alternative approach to thinking about theorems of this sort is due to \citet{Pitowsky}.  I should note, though, that the version of Bell's theorem I will present in section \ref{bell} of the present paper is due to \citet{CHSH} and is somewhat different than the versions discussed by Jarrett and Malament (which are in the tradition of \citet{CH}) or Pitowsky.  I will make some contact between the present discussion and Jarrett-Malament tradition of presenting Bell-type theorems in the final section of the paper.}

In this paper, I discuss a geometrical model of EPR-Bohm experiments that appears, at least at first blush, to get around Bell-type theorems.\footnote{The model I discuss here is inspired by, but not the same as, models discussed by Joy Christian. (See \citet{Christian1, Christian2, Christian3, Christian4, Christian5, Christian6, Christian7, Christian8} and other papers available online; these are edited and collected with some new material in \citet{ChristianBook}.)  To my mind, at least, the present model is considerably simpler than Christian's, and it avoids certain technical complications that his models encounter.  (For more on these technical issues, see in particular \citet{Moldoveanu} and references therein.)  That said, although my engagement with Christian's work will be from an oblique angle, what I say applies equally well to the models he discusses.  I should also note that since the present paper first appeared online, Christian as posted a reply \citep{Christian-Reply}.  However, his reply is not responsive to the issues raised here and for this reason I do not engage with it.}  The model offers an explicit representation of the states of the particles in an EPR-Bohm experiment.  It is manifestly local and fully deterministic.  And it appears to yield the correct measurement outcomes, as predicted by quantum mechanics.  As one would expect given what I have just said, the model I will describe differs from the kind of system Bell seemed to have in mind in several important ways---ways, I might add, that are physically motivated by the EPR-Bohm experiment itself.\footnote{In personal correspondence and elsewhere, Christian has criticized the model presented here as \emph{ad hoc} and unrealistic.  This charge is entirely just.  But I should be clear that the model is not meant to be a realistic contender to explain Bell-violating correlations in quantum mechanics; rather, it is a simple toy model with certain suggestive formal properties that helps clarify the explanatory demands that a local, deterministic model of the EPR-Bohm experiment would have to meet.  The argument here is not that the present model fails and thus Christian's (unquestionably different) model also fails; it's that a careful analysis of the present simple model reveals something important about what any successful model must do.  I then show that no model of a certain quite general form can do this.}

Unfortunately, the model is not ultimately satisfactory, in ways I will be able to spell out clearly later in the paper.  But it seems to me that it fails in an interesting way.  The model draws attention to just what criteria a local, deterministic hidden variable model would need to satisfy in order to accurately reproduce the results of an EPR-Bohm experiment.  The particular way in which the present model fails to meet those criteria speaks to the generality of Bell-type theorems.  Taking the model as a guide, I will formulate and prove a minor proposition to the effect that a class of hidden variable models similar in a certain respect to the one I describe cannot reproduce the EPR-Bohm measurement results, properly understood.  The proposition follows as a corollary of a version of Bell's theorem.  This means that models of the type I will describe are already ruled out by Bell-type considerations, despite the apparent differences between the model I will discuss and the kind of hidden variable models with which Bell-type theorems seem to be concerned.\footnote{The character of the response offered here is related to a brief criticism of Christian's proposal by \citet{Grangier}, although the details are quite different.}

It is worth emphasizing up front that I do not take the main proposition of this paper to be a significant new contribution.  Indeed, there is a strong sense in which the result, though it might appear to be a generalization of standard Bell-type theorems, simply \emph{is} Bell's theorem, or even a slight weakening of it.  The goal of this paper is instead to show why Bell-type theorems really do have the scope and generality standardly ascribed to them, by explaining how a particular strategy for getting around Bell-type theorems goes wrong.  The main proposition, then, is just an attempt to re-phrase Bell's theorem in a way that makes the connection between Bell-type hidden variable models and these alternative, allegedly local and deterministic ``generalized'' hidden variable models fully transparent.\footnote{Does this point need to be made in print?  I am sympathetic with readers who might think that it does not.  But the work by Christian alluded to above, and the significant controversy it has spawned in some circles (see for instance the discussion at Scott Aaronson's blog, http://www.scottaaronson.com/blog/?p=1028), suggests that there remains considerable confusion regarding just what Bell-type theorems tell us, even among serious physicists.}

The paper will proceed as follows.  I will begin by describing the EPR-Bohm setup and stating the version of Bell's theorem that I will focus on here.  Then I will present the model that will be the focus of the present discussion.  I will show that it is local and deterministic, but that it nonetheless seems to reproduce the quantum mechanical predictions for the EPR-Bohm experiment.  In the penultimate section, I will describe what goes wrong with the model and then formulate and prove the no-go result described above.  I will conclude with some remarks concerning the generality of Bell-type theorems.

\section{The EPR-Bohm experiment and a Bell-type Theorem}\label{bell}

The version of Bell's theorem I will discuss here is based on an experimental configuration known as the EPR-Bohm setup, which involves two spin 1/2 particles, $A$ and $B$, traveling in opposite directions.  One assumes that the pair initially has vanishing total spin, which means that the quantum mechanical spin state for the system, $\ket{\Psi}$, is an entangled superposition often called the ``singlet state.''  A natural way to express the singlet state is in terms of the eigenvectors of the spin operator about an arbitrary unit vector $\mathbf{n}$.  The spin operator about $\mathbf{n}$ is given by $\boldsymbol{\sigma}\cdot\mathbf{n}$, where $\boldsymbol{\sigma}$ is the Pauli spin ``vector,'' defined by $\boldsymbol{\sigma}=(\sigma_x,\sigma_y,\sigma_z)$, with $\sigma_x,\sigma_y,\sigma_z$ the Pauli spin matrices.\footnote{The Pauli spin vector is actually a map from unit vectors in three dimensional Euclidean space to the space of operators on the two-dimensional Hilbert space representing spin states.  But the abuse of notation should be harmless, since the inner product is shorthand for the obvious thing: $\boldsymbol{\sigma}\cdot\mathbf{n}=\sigma_xn_x+\sigma_yn_y+\sigma_zn_z$.  Note that we are working in units where $\hbar=2$.  }
These eigenvectors can be written in the standard Dirac notation so that $\boldsymbol{\sigma}\cdot\mathbf{n}\ket{\mathbf{n},\pm}=\pm\ket{\mathbf{n},\pm}$.
The singlet state of the two-particle system can then be expressed by
\[\ket{\Psi}=\frac{1}{\sqrt{2}}\left(\ket{\mathbf{n},+}_A\otimes\ket{\mathbf{n},-}_B-\ket{\mathbf{n},-}_A\otimes\ket{\mathbf{n},+}_B\right),\]
where the $A$ and $B$ subscripts indicate membership in the Hilbert spaces associated with particles $A$ and $B$ respectively.

The experiment consists of two measurements, one on each particle, performed at remote locations by two observers, Alice and Bob.  Alice and Bob each choose a unit vector, $\mathbf{a}$ and $\mathbf{b}$ respectively, and then perform measurements of the spin of their assigned particle about the chosen vector (suppose Alice is measuring particle $A$ and Bob is measuring particle $B$).  Simple calculations yield the following quantum mechanical predictions for the expected outcomes of these experiments.  The expectation values for Alice and Bob's individual experimental outcomes are,
\begin{align}
\mathcal{E}_{q.m.}(A(\mathbf{a}))&=\bra{\Psi}(\boldsymbol{\sigma}\cdot\mathbf{a})_A\otimes\mathbb{I}_B\ket{\Psi}=0\\
\mathcal{E}_{q.m.}(B(\mathbf{b}))&=\bra{\Psi}\mathbb{I}_A\otimes(\boldsymbol{\sigma}\cdot\mathbf{b})_B\ket{\Psi}=0.
\end{align}
Meanwhile, the expectation value of the (tensor) product of the two observables $(\boldsymbol{\sigma}\cdot\mathbf{a})_A\otimes(\boldsymbol{\sigma}\cdot\mathbf{b})_B$, a measure of the correlation between the two measurements, is given by
\begin{equation}\label{qmev}
\mathcal{E}_{q.m.}(A(\mathbf{a}),B(\mathbf{b}))=\bra{\Psi}(\boldsymbol{\sigma}\cdot\mathbf{a})_A\otimes(\boldsymbol{\sigma}\cdot\mathbf{b})_B\ket{\Psi}= -\mathbf{a}\cdot\mathbf{b}.
\end{equation}
To interpret this expectation value, it is useful to note in particular that if Alice and Bob choose to perform their measurements about orthogonal vectors, their outcomes should not exhibit any correlation at all (they are equally likely to get the same results as opposite results).  Conversely, if Alice and Bob choose the same vector, their measurement results should be perfectly anti-correlated (any time Alice yields +1, Bob necessarily will yield -1, and vice versa).  These predictions are well-corroborated by experiment.

Bell-type theorems are intended to rule out the possibility that the results of EPR-Bohm experiments can be recovered from a model that somehow represents the ``complete'' state of the two-particle system in such a way that the results of Alice and Bob's measurements can be understood to be local and deterministic.  The strategy is to postulate that there is an unknown (hidden) variable, $\lambda$, such that, were one to specify the value of $\lambda$, the results of Alice and Bob's respective measurements would be (a) wholly determined by $\lambda$ and the choice of his or her measurement vector (call this ``determinism''), and (b) independent of the other experimenter's choice of measurement vector (call this ``locality'').  One then tries to show that no such model could correctly reproduce the predictions of quantum mechanics.

This idea can be made more precise as follows.  We postulate that associated with each of Alice's and Bob's measuring devices are ``observables'' $A$ and $B$, where by observable we mean simply maps $A,B:\mathbb{S}^2\times\Lambda\rightarrow\{-1,1\}$ that take as input (1) the unit vector about which Alice or Bob is measuring, and (2) an element $\lambda$ of the space $\Lambda$ of complete states of the two particle system.\footnote{The set $\mathbb{S}^2\subseteq\mathbb{R}^3$ is to be understood as the 2-sphere, i.e., the set of unit vectors in $\mathbb{R}^3$.} The specification $\lambda$ of the complete state of the two-particle system is what we interpret as the hidden variable.  The possible measurement results, meanwhile, are represented by the integers $\pm 1$, just as in the quantum mechanical case (where the possible measurement results correspond to the eigenvalues of the relevant observable).

Given such a pair of observables $A$ and $B$, we can define the expectation values for Alice and Bob's measurements by postulating some probability density $\rho:\Lambda\rightarrow [0,1]$, intended to reflect our ignorance of the actual value of the hidden variable.  The expectation values for Alice and Bob's individual measurements will then be given by:\footnote{In the following expressions, I am writing integrals over $\Lambda$ without being fully specific about what $\Lambda$ looks like, with the possible consequence that the interpretation of the expressions is ambiguous.  It turns out, though, that Bell-type results are based on features of integrals that are so basic that it does not matter what kind of integral is being written.}
\begin{align*}
\mathcal{E}_{h.v.}(A(\mathbf{a}))&=\int_{\Lambda}A(\mathbf{a},\lambda)\rho(\lambda)d\lambda\\
\mathcal{E}_{h.v.}(B(\mathbf{b}))&=\int_{\Lambda}B(\mathbf{b},\lambda)\rho(\lambda)d\lambda.
\end{align*}
The version of Bell's theorem to be stated presently will be formulated as a constraint on the expectation value of the product of $A$ and $B$, \[\mathcal{E}_{h.v.}(A(\mathbf{a})B(\mathbf{b}))=\int_{\Lambda}A(\mathbf{a},\lambda)B(\mathbf{b},\lambda)\rho(\lambda)d\lambda.\]
As in the quantum mechanical case, this expectation value is understood as a measure of the correlation between Alice and Bob's measurements.

This discussion suggests the following definitions.
\begin{defn}\singlespacing \label{hvm}A (local, deterministic) \emph{hidden variable model} of the EPR-Bohm experiment is an ordered quadruple $(\Lambda, A, B,\rho)$, where $\Lambda$ is the set of complete states of the system, $A,B:\mathbb{S}^2\times\Lambda\rightarrow \{-1,1\}$ are maps from measurement vectors and complete state specifications to measurement outcomes for Alice and Bob respectively, and $\rho:\Lambda\rightarrow[0,1]$ is a probability density function on the space of complete states.\end{defn}
Note that determinism is built into this definition, in the sense that the maps $A$ and $B$ are required to be well-defined as functions, which means that there is a unique measurement result associated with each specification of a measurement vector and a complete state.\footnote{At times I will continue to refer to a ``local, deterministic hidden variable model,'' but the ``local'' and ``deterministic'' modifiers will be simply for emphasis.  As I will understand it in the rest of this paper, a hidden variable model is automatically local and deterministic.}  Similarly locality is encoded in the requirement that the maps $A$ and $B$ are functions of a single measurement vector.  (A non-local model, then, would be one on which $A$ and $B$ were allowed to depend on both measurement vectors.)

Given these definitions, we can now state the Bell-type theorem that we will focus on in the next section.  It is essentially the Clauser-Horne-Shimony-Holt theorem.
\begin{thm}[\citet{CHSH}]\label{chsh}\singlespacing
Let $(\Lambda, A, B,\rho)$ be a local, deterministic hidden variable model of the EPR-Bohm experiment. Then for any choices $\mathbf{a}$ and $\mathbf{a}'$ for Alice's detector setting, and any choices $\mathbf{b}$ and $\mathbf{b}'$ for Bob's detector setting, the expectation values for the products of Alice and Bob's measurements must satisfy the Clauser-Horne-Shimony-Holt inequality,
\begin{equation}
|\mathcal{E}_{h.v.}(A(\mathbf{a})B(\mathbf{b}))-\mathcal{E}_{h.v.}(A(\mathbf{a})B(\mathbf{b}'))+\mathcal{E}_{h.v.}(A(\mathbf{a}')B(\mathbf{b})) + \mathcal{E}_{h.v.}(A(\mathbf{a}')B(\mathbf{b}'))|\leq 2.
\end{equation}
\end{thm}
It is easy to verify that there are choices for $\mathbf{a}$, $\mathbf{a}'$, $\mathbf{b}$, and $\mathbf{b}'$ such that the quantum mechanical expectation values defined in Eq. \eqref{qmev} violate this inequality, i.e., choices such that
\[
|\mathcal{E}_{q.m.}(A(\mathbf{a})B(\mathbf{b}))-\mathcal{E}_{q.m.}(A(\mathbf{a})B(\mathbf{b}'))+\mathcal{E}_{q.m.}(A(\mathbf{a}')B(\mathbf{b})) + \mathcal{E}_{q.m.}(A(\mathbf{a}')B(\mathbf{b}'))|> 2.
\]
This result yields an immediate corollary of Thm. \ref{chsh}.
\begin{cor}\label{nogo1}\singlespacing
There does not exist a local, deterministic hidden variable model of the EPR-Bohm experiment that reproduces the quantum mechanical expectation values.
\end{cor}

\section{A local, deterministic model after all?}\label{uhoh}

As promised above, I will now explicitly exhibit what might seem to be a local, deterministic hidden variable model of the EPR-Bohm experiment.  It will consist of a space $\Lambda$ of complete states, along with a pair of local, deterministic observables $A(\mathbf{a},\lambda)$ and $B(\mathbf{b},\lambda)$ that, taken together, yield the correct quantum mechanical expectation values.  Note that this means they will maximally violate the Clauser-Horne-Shimony-Holt inequality for the same choices of $\mathbf{a}$, $\mathbf{a}'$, $\mathbf{b}$, and $\mathbf{b}'$ as the quantum mechanical observables.  As will become clear, this model is not a hidden variable model in the strict sense of definition \ref{hvm}.  But one might reasonably argue that it has all of the virtues that one would hope a (generalized) hidden variable model would have.

Before I say what the model is, let me motivate it a little.  The experiment whose outcomes we are trying to reproduce involves spin measurements.  Quantum mechanical spin does not have a direct classical analogue, but the spin operators satisfy the same algebraic relations as rotations in three dimensional Euclidean space.  This is suggestive: one might take it to imply that the right way of thinking about spin is as some sort of rotation.  Following this idea, one might reason that when an experimenter measures a particle to be spin up or spin down about a given vector, she is actually measuring the orientation of the body's rotation.

Suppose this idea is right and the experimenters are attempting to determine the orientation of each body's rotation about freely chosen vectors $\mathbf{a}$ and $\mathbf{b}$.  How should the experimenters represent their measurement results?  Clearly, one way of doing so would be to assign the numbers $\pm1$ to the measurement results, where, say, $+1$ is understood to correspond to spin up, which would be a counterclockwise rotation about the measurement vector (relative to a right-hand rule), and $-1$ would correspond to spin down and clockwise rotation.  This is how the quantum mechanical expectation values come out; it is also how possible measurement results are represented in Bell-type theorems.  But one might worry that this is a poor choice.  Rotations in three dimensional space have more complicated algebraic and topological structure than the set $\{-1,1\}\subset\mathbb{R}$.  When one represents the results of an EPR-Bohm experiment by $\pm1$, this structure is lost.  One might argue that it is no wonder that a hidden variable model of the type Bell describes cannot adequately represent the results of an EPR-Bohm experiment, since Bell effectively assumes a representational scheme that does not have the expressiveness necessary to capture the physics of the experiment.\footnote{There are several reasons why a reader might balk at this point.  I admit that the argument given here is problematic; indeed, I will attempt to say, as precisely as possible, what goes wrong with this argument in the next section.  In the meantime, I encourage a reader who anticipates the problems even at this stage to hold back her objections to see where the discussion goes.  This is a productive garden path to wander down, even if one can already see that it ends in a coal pit.}

There are several ways to represent rotations (or more precisely, instantaneous states of rotation of a body) in three dimensional space that do respect the appropriate algebraic and topological properties of physical rotations.  A particularly natural choice is to use antisymmetric rank 2 tensors.  To see how these are a candidate for representing a body's state of rotation, consider the following.  Pick a plane and consider two linearly independent covectors in that plane.  Call them $\xi_a$ and $\eta_a$.\footnote{At this point, I am changing notations, since the model I will presently describe is most naturally expressed using tensor fields defined on a manifold.  Here and throughout I will use the abstract index notation explained in \citet{Penrose+Rindler} and \citet{MalamentGR}.  For present purposes, one can safely think of these as the counting indices of coordinate notation, with an Einstein summation convention assumed.  I will use the following translation manual for vectors previously discussed: I will now use $\alpha^a$ instead of $\mathbf{a}$ to represent the vector about which Alice chooses to make her measurement, and $\beta^a$ instead of $\mathbf{b}$ to represent the vector about which Bob chooses to make his measurement.  These vectors still live in $\mathbb{S}^2$.}  Now note that there are two orientations of rotations within your chosen plane, clockwise and counterclockwise.  These can be thought of as corresponding to the two ways of ordering $\xi_a$ and $\eta_a$: one of the two orientations corresponds to a rotation that begins with $\xi_a$ and sweeps in the direction of $\eta_a$, and the other corresponds to a rotation that begins with $\eta_a$ and sweeps in the direction of $\xi_a$.  (Think, for instance, of a clock face.  Clockwise rotation corresponds to the minute hand starting at 12 and sweeping towards 1; counterclockwise orientation corresponds to the minute hand starting at 1 and sweeping towards 12.)  These two ways of ordering $\xi_a$ and $\eta_a$ can be (sufficiently uniquely) represented using a pair of related antisymmetric rank 2 tensors, $\overset{1}{\omega}_{ab}=\xi_{[a}\eta_{b]}=\frac{1}{2}(\xi_{a}\eta_{b}-\xi_{b}\eta_{a})$ and $\overset{2}{\omega}_{ab}=\eta_{[a}\xi_{b]}=-\overset{1}{\omega}_{ab}$.  Indeed, in three dimensions, any antisymmetric rank 2 tensor can be represented as the antisymmetric product of two vectors, and so any antisymmetric rank 2 tensor represents a rotation in just this way, as an ordering of two vectors in a plane.  This means that to associate an antisymmetric rank 2 tensor with a body is to represent the body as having both a plane of rotation (or equivalently, an axis of rotation) and an orientation of rotation within that plane.\footnote{What does it mean to say that representing states of rotation as antisymmetric rank 2 tensors respects the ``algebraic and topological properties'' of rotations in three space?  The space of constant antisymmetric rank 2 tensors on three dimensional Euclidean space (understood as a vector space) is itself a three dimensional vector space over $\mathbb{R}$ (i.e., it is closed under addition and scalar multiplication) and moreover forms a Lie algebra with the Lie bracket defined by, for any two antisymmetric rank 2 tensors $\overset{1}{\omega}_{ab}$ and $\overset{2}{\omega}_{ab}$, $[\overset{1}{\omega},\overset{2}{\omega}]_{ab}=\overset{1}{\omega}_{an}\overset{2}{\omega}{}^n{}_{b}-\overset{2}{\omega}_{an}\overset{1}{\omega}{}^n{}_{b}$.  A short calculation shows that this Lie algebra is none other than $\mathfrak{so}_3$, the Lie algebra associated with the Lie group of rotations of three dimensional vectors, $SO(3)$.  The elements of $SO(3)$, meanwhile, can be represented by the length preserving maps between vectors in $\mathbb{R}^3$ that are standardly identified with rotations.  Note that since the Lie algebra $\mathfrak{so}_3$ is isomorphic to the Lie algebra $\mathfrak{su}_2$, the vector space generated by the Pauli spin matrices, which are the operators associated with spin for a spin 1/2 system, is also a representation of $\mathfrak{so}_3$.}

We are thus led to the following suggestion.  If one wants to adequately represent the rotation of a body about a given axis, taking full account of the algebraic and topological properties exhibited by the group of rotations in three dimensional space, one should take the observables associated with the EPR-Bohm measurement outcomes to be antisymmetric rank 2 tensors.\footnote{These are not ``observables'' in anything like the standard sense. They might be better characterized as complete representations of the system, and perhaps the apparatus, after measurement.  I will at times use the term ``observable'' in this non-standard sense since these ``observables'' are intended to play a similar formal role to the observables in a Bell-type hidden variable model.}$^,$\footnote{There are other ways one might do this, too.  Christian, for instance, proposes that one represent rotations using bivectors in a Clifford algebra.  There are various reasons to think that that proposal is equally well-motivated, or even better motivated, than the one I make here.  But ultimately such distinctions will not matter.  See the next section for more on this point.}  This is particularly important, one might add, since we are ultimately interested in the correlation between the two systems.  Hence we want to be careful to capture the full structure of what is being observed, since one might expect that the algebraic properties of rotations are important in understanding how the outcomes of two measurements of rotation will relate to one another.  Of course, taking such objects as observables moves one away from a hidden variable model as defined in section \ref{bell}.  But that need not be a problem, especially if it turns out that all one needs to do to find a local, deterministic interpretation of quantum mechanics is to expand the mathematical playing field slightly.

Given this background, the model I propose goes as follows.  Our starting point will be the manifold $\mathbb{R}^3$, endowed with a (flat) Euclidean metric $h_{ab}$.  The space of constant vector fields on the manifold is isomorphic to the tangent space at any one point (by parallel transport); I will conflate these two spaces and refer to both simply as $\mathbb{R}^3$.  One can likewise define a space of (constant) antisymmetric rank 2 tensors by taking the antisymmetric product of the elements of $\mathbb{R}^3$, as described above.  I will call this space of tensors $\mathbb{R}^3\wedge\mathbb{R}^3$.  Note that there are precisely two volume elements (normalized antisymmetric rank 3 tensor fields) on the manifold.  Call these $\epsilon_{abc}$ and $-\epsilon_{abc}$.  We will fix a sign convention up front and take $\epsilon_{abc}$ to represent the right handed volume element.\footnote{A volume element can be understood to give an orientation to the entire three dimensional manifold in much the same way that antisymmetric rank 2 tensors give an orientation to a plane, by specifying the relative order of any three linearly independent vectors.  For more on volume elements, see \citet{MalamentGR}.}

The hidden variable space for the model will be $\Lambda=\{1,-1\}$.  We can then define two observables $A,B:\mathbb{S}^2\times\Lambda\rightarrow\mathbb{R}^3\wedge\mathbb{R}^3$ as follows:
\begin{align*}
A_{bc}(\alpha^a,\lambda)&=\frac{\lambda}{\sqrt{2}}\epsilon_{abc}\alpha^a\\
B_{bc}(\beta^a,\lambda)&=-\frac{\lambda}{\sqrt{2}}\epsilon_{abc}\beta^a
\end{align*}
Note that these observables are both antisymmetric rank 2 tensors, as we wanted.  They can be understood to represent rotations about the vectors $\alpha^a$ and $\beta^a$, as can be seen by noting that $A_{bc}(\alpha^a,\lambda)\epsilon^{abc}\propto \alpha^a$ and $B_{bc}(\beta^a,\lambda)\epsilon^{abc}\propto \beta^a$.  Note, too, that for a particular choice of $\lambda$, and for measurements about parallel vectors, we find \[A_{bc}(\xi^a,\lambda)=-B_{bc}(\xi^a,\lambda).\]  This means the two measurement results are always anti-correlated, as expected from the quantum mechanical case.  Finally, $A$ and $B$ are explicitly deterministic, since given a value for the hidden variable, and given a choice of measurement vector, the observables take unique, determinate values.  And they are local, in the sense that the value of the observable $A$ is independent of $\beta^a$, and $B$ is independent of $\alpha^a$.

We assume an isotropic probability density over the space of hidden variables.  This density is simply a map $\rho:\Lambda\rightarrow[0,1]$ that assigns equal probability to both cases.  With this choice, the observables can easily be shown to yield the right single-side expectation values.\footnote{Actually, there is a slight abuse here: expectation values are usually defined to be real numbers; the ``expectation values'' I define presently are actually rank 2 tensors (specifically in this case the zero tensor).  However, the interpretation of these generalized expectation values is unambiguous, and they clearly have the desired physical significance.}  We have for both $A$ and $B$,
\begin{align}
\mathcal{E}_{h.v.}(A(\alpha^a))&=\frac{1}{2}(\frac{1}{\sqrt{2}}\epsilon_{abc}\alpha^a-\frac{1}{\sqrt{2}}\epsilon_{abc}\alpha^a)=\mathbf{0}\\
\mathcal{E}_{h.v.}(B(\beta^a))&=-\frac{1}{2}(\frac{1}{\sqrt{2}}\epsilon_{abc}\beta^a-\frac{1}{\sqrt{2}}\epsilon_{abc}\beta^a)=\mathbf{0}.
\end{align}
Hence, on average, neither body is rotating about \emph{any} direction.  This result is consistent with the quantum mechanical expectation values and with experiment.

Finally, we want to consider the expectation value of the product of the measurement results.  This is where the algebraic properties of rotations become important, since it is at this point that one is trying to establish a relationship between two correlated, rotating bodies.  But now consider the product of the outcomes:\[A_{bc}(\alpha^a,\lambda)B^{bc}(\beta^a,\lambda)=-\frac{\lambda^2}{2}\alpha^n\epsilon_{nbc}\beta_m\epsilon^{mbc}=-\alpha^n\beta_n.\]
The associated expectation value, then, is just \begin{equation}\label{jcje}
\mathcal{E}_{h.v.}(A(\alpha^a)B(\beta^a))=-\alpha^a\beta_a(\frac{1}{2}+\frac{1}{2})=-\alpha^a\beta_a,
\end{equation} again just as in the quantum mechanical case.\footnote{It is at this stage that the model I present diverges most substantially from Christian's.  In his proposal, the product expectation value is calculated using the geometric product of two Clifford algebra-valued observables.  In mine, it is the inner product of two antisymmetric rank 2 tensor fields.  Antisymmetric rank 2 tensor fields can generally be identified with bivectors in a Clifford algebra, but the products in the two cases are entirely different.  It turns out, though, that this distinction just does not matter.  Neither product is tracking the right experimental information, and indeed, as I will presently show, no such product \emph{could} track the right experimental information and still yield a result that conflicts with Bell-type theorems.}$^,$\footnote{This expectation value is a real expectation value, in the sense that it is real-valued, unlike the two single sided expectation values defined above.}   With this final result, it would appear that we have written down a model of the EPR-Bohm experiment.  And more, the product expectation value associated with this model will violate the Clauser-Horne-Shimony-Holt inequality in exactly the same way as quantum mechanics---after all, it has just the same functional form.\footnote{The present point will become clear in what follows, but it perhaps worth emphasizing here as well: the conclusion stated in the text above is \emph{false}.  That is, the model I have described in this section, and particularly the expressions for the product of the measurement results and its expectation value, do \emph{not} reproduce the predictions of quantum mechanics.  The goal of the remainder of the paper will be to articulate as clearly as I can where the problems lie.}

\section{A diagnosis and a no-go result}\label{nogo}

What are we to make of the results of sections \ref{bell} and \ref{uhoh}?  One might argue that they do not actually conflict.  Instead, the model just presented reveals that the notion of hidden variable model given by definition \ref{hvm} is too narrow.  There, a hidden variable model was defined to have two observables that were integer valued---specifically, valued by elements of the set $\{-1,1\}$.  Rotations in three dimensions, meanwhile, are a complicated business.  The correlations exhibited by two rotating bodies cannot be captured in terms of a two element set.  One could conclude that it is necessary to consider ``observables'' whose possible values reflect the full structure of rotations, as antisymmetric rank 2 tensors do.  And indeed, if one does permit such observables, it would appear that one can construct a local, deterministic (generalized) hidden variable model of the EPR-Bohm experiment after all.

On this view, the model just described is not a counterexample to Bell-type theorems \emph{per se}, but it does undermine their foundational importance.  Bell-type theorems were of interest because they were understood to exclude a particular, appealing sort of hidden variable theory.  If it were to turn out that they only applied to a limited number of those theories---ones with integer-valued observables---the standard moral of Bell-type arguments would be severely muted.  There would be hope for finding a local, deterministic (generalized) hidden variable theory for quantum mechanics after all.

Unfortunately, this argument is too fast.  The model I have described is not acceptable.  The problem concerns what, exactly, a (generalized) hidden variable model needs to reproduce.  Specifically, Eq. \eqref{jcje} does not correctly track the relevant measure of correlation between Alice and Bob's measurement results.

To see the point, consider the following.  When Alice and Bob actually run their experiments, they are perfectly entitled to record their measurement results however they choose.  They can even record those results as integers $\pm1$, so long as they do so systematically.  If a model like the one given above is correct, then what Alice and Bob would be doing would amount to applying some map $P$ from the codomain of their observables to the set $\{-1,1\}$.  Of course, there is a danger that representing their results in this way could turn out to be a poor choice.  One might worry that such a recording mechanism would lose information about their measurement results.  If the thing they are observing has more structure than the set $\{-1,1\}$, then Alice and Bob would risk introducing confusing patterns into their data.  Indeed, we already know that the data from EPR-Bohm experiments, standardly represented as sequences of $\pm1$, do exhibit correlations that are difficult to interpret.  But here is the crucial point.  Whatever else is the case, the troublesome correlations that are present in EPR-Bohm data arise when the data is recorded as sequences of $\pm1$.  And so it is \emph{these} correlations that a (generalized) hidden variable model needs to reproduce in order to be satisfactory.  In other words, to show that a (generalized) hidden variable model correctly reproduces the EPR-Bohm experimental results, one needs to show that the model recovers the standard quantum mechanical expectation values \emph{after} we project the generalized observables onto $\{-1,1\}$.

So in order to determine whether the model presented above is satisfactory, we need to find a way for Alice and Bob to systematically project the relevant elements of $\mathbb{R}^3\wedge\mathbb{R}^3$ onto $\{-1,1\}$.   There is a principled way to do this.  It is to take the integer to indicate the orientation of the rotation about the measurement vector (so, $+1$ might record counterclockwise rotation, $-1$ clockwise).  In the present case, this representational scheme can be expressed by a family of maps defined as follows.  Let $\xi^a$ be a choice of measurement vector.  Then there is an associated projection map $P_{\xi}:(\overline{\mathbb{R}^3\wedge\mathbb{R}^3})_{\xi}\rightarrow \{-1,1\}$,\footnote{To be clear, the Greek index on $P$ is not a tensor index---rather, it indicates which measurement vector one is projecting relative to.  In what follows, I will at times suppress this index to talk generally about a projection relative to some unspecified measurement vector.} where $(\overline{\mathbb{R}^3\wedge\mathbb{R}^3})_{\xi}\subset \mathbb{R}^3\wedge\mathbb{R}^3$ is the collection of antisymmetric rank 2 tensor fields $\omega_{ab}$ that (1) are orthogonal to $\xi^a$ (i.e., $\omega_{ab}\xi^a=\mathbf{0}$) and (2) satisfy the normalization condition $\omega_{ab}\omega^{ab}=1$.\footnote{It is worth observing that the observables defined above, evaluated for a given measurement vector, are in this set, relative to that choice of measurement vector.}  The map associated with each measurement vector turns out to be unique (up to sign).  For a given unit vector $\xi^a$, and for any antisymmetric rank 2 tensor field $X_{ab}$ in $(\overline{\mathbb{R}^3\wedge\mathbb{R}^3})_{\xi}$, the map can be defined as,
\[
P_{\xi}(X_{ab})=\frac{1}{\sqrt{2}}X_{ab}\epsilon^{nab}\xi_n.
\]
Note that projecting in this way may be construed as a loss of structure, insofar as $\{-1,1\}$ does not have the algebraic properties of rotations.  But it is as good a way as any to represent an antisymmetric rank 2 tensor as an integer, which is what we need to do to adequately evaluate the model.

What, then, does the model predict for Alice and Bob's measurements, recorded using these projection maps?  For Alice, for any given choice of measurement vector $\alpha^a$, the answer is given by,
\[
P_{\alpha}(A_{ab}(\alpha^a,\lambda))=\frac{1}{\sqrt{2}}(\frac{\lambda}{\sqrt{2}}\epsilon_{mab}\alpha^m)\epsilon^{nab}\alpha_n=\lambda.
\]
For Bob, meanwhile, for any choice of $\beta^a$, we find,
\[
P_{\beta}(B_{ab}(\beta^a,\lambda))=\frac{1}{\sqrt{2}}(-\frac{\lambda}{\sqrt{2}}\epsilon_{mab}\beta^m)\epsilon^{nab}\beta_n=-\lambda.
\]
By definition $\lambda=\pm 1$, so these results are correctly valued in the set $\{-1,1\}$.  Since we have assumed a uniform density function, it follows that
\begin{align*}
\mathcal{E}_{h.v.}(P_{\alpha}(A_{ab}(\alpha^a))&=\frac{1}{2}-\frac{1}{2}=0\\
\mathcal{E}_{h.v.}(P_{\beta}(B_{ab}(\beta^a))&=-\frac{1}{2}+\frac{1}{2}=0,
\end{align*}
which means that Alice and Bob should be getting $+1$ and $-1$ about half the time each.  This is in agreement with the quantum mechanical case.

But something is awry.  Suppose that Alice and Bob pick their respective vectors $\alpha^a$ and $\beta^a$ and then perform successive measurements with these measurement vectors, recording the results as $\pm1$ as they go.  After many trials, Alice and Bob should have long lists of numbers.  For Alice and Bob individually, if you took the average value of each of their results, you would find it tending to 0. This is what it means to say that the individual expectation values are zero.  But now, what about the expectation value of the product of these?  This value is supposed to be a prediction for the average of the product of Alice and Bob's measurement results for each trial, recorded as integers.  So, if Alice wrote down $+1$ for her first trial, and Bob wrote down $+1$, the product of their outcomes for that trial would be $+1$.  If for the next trial, Alice wrote down $+1$ and Bob wrote down $-1$, the product would be $-1$.  And so on.  The quantum mechanical expectation value for the product of the measurement results, $\mathcal{E}_{q.m.}(AB)$, amounts to a prediction that, over the long run, the sequence of values of $\pm1$ calculated by multiplying Alice and Bob's individual results for each trial will have an average value of $-\alpha^a\beta_a$.  This is the prediction that is confirmed by the experimental evidence, that violates the Clauser-Horne-Shimony-Holt inequality, and that would have to be accounted for by a viable local, deterministic (generalized) hidden variable theory.

Does the model presented in section \ref{uhoh} accomplish this task?  Decidedly not.  Consider what happens on this model if Alice and Bob choose to record their results using $\pm1$, via the projection maps defined above.  One immediately sees that for \emph{any} choices of $\alpha^a$ and $\beta^a$, if you take the product of Alice's result and Bob's result for a particular trial, you always get the same number.  Specifically,
\[
P_{\alpha}(A_{ab}(\alpha^a,\lambda)) P_{\beta}(B_{ab}(\beta^a,\lambda))=-\lambda^2=-1.
\]
It follows that the relevant product expectation value, the expectation of the product of the results after projection, is,
\[
\mathcal{E}_{h.v.}(P_{\alpha}(A_{ab}(\alpha^a)) P_{\beta}(B_{ab}(\beta^a)))=\frac{1}{2}(-1+-1)=-1.
\]
This is not consistent with the quantum mechanical case, nor with experiment.  (It also fails to violate the Clauser-Horne-Shimony-Holt inequality.)\footnote{We can now see that the argument in the previous section was a straightforward paralogism.  It goes, (1) quantum mechanics predicts that the average value of the products of two sequences of integers will be the inner product of two vectors; (2) this alternative model predicts that the average value of the inner products of two sequences of tensors will be the inner product of two vectors; therefore (3) the alternative model reproduces the predictions of quantum mechanics.  Since (1) and (2) make reference to two different sequences, (3) does not follow.}

As I have already suggested, the problem stems from the product expectation value defined in Eq. \eqref{jcje}.  There we appeared to show that the product expectation value for Alice and Bob's measurements was $-\alpha^a\beta_a$, as expected from quantum mechanics.  But we now see that we were comparing apples to oranges.  The calculation in Eq. \eqref{jcje} is a formal trick, something that bears a passing resemblance to the expectation value we want to calculate, but which is actually irrelevant to the problem we were initially interested in.  Eq. \eqref{jcje} does give \emph{some} measure of the overlap between Alice and Bob's results, but not the measure that we were trying to reproduce.

The considerations just given lead to a more general result.  To state it, we begin by defining a new, broader notion of hidden variable model.\footnote{Calling a model of the sort I define above ``generalized'' is admittedly tendentious.  There is a strong sense in which this sort of model is a \emph{specialization} rather than a generalization of Bell's notion of a hidden variable model, since (as will be clear in the proof of Corollary \ref{nogo2}) it implicitly includes all of the pieces of a Bell-type hidden variable model, plus some additional (unnecessary) machinery.  Nonetheless I am using the expression, since it captures the intended spirit of the (failed) model described in the body of the paper.}
\begin{defn}\singlespacing \label{hvmg}A (local, deterministic) \emph{generalized hidden variable model} of the EPR-Bohm experiment is an ordered sextuple $(\Lambda, X, A, B,\rho, \mathcal{P})$, where $\Lambda$ is a space of complete states of the system, $X$ is some set of possible states of the system and apparatus after measurement, $A,B:\mathbb{S}^2\times\Lambda\rightarrow X$ are maps from measurement vectors and complete state specifications to states of the system and apparatus after measurement for Alice and Bob respectively, $\rho:\Lambda\rightarrow[0,1]$ is a probability density function on the space of complete states, and $\mathcal{P}=\{P_{i}:X\rightarrow\{-1,1\}\}_{i\in\mathbb{S}^2}$ is a family of maps that provide a way to represent the elements of $X$ as numbers $\pm 1$ for each choice of measurement vector.\footnote{In principle, the family of maps could also be parameterized by values of $\Lambda$ and everything below would go through with only minor modifications.  I do not explicitly treat this possible generalization, however, because its significance is unclear: if one thinks of the projection as an operation that an experimenter performs, then it is hard to see how that operation could depend on a variable whose value the experimenter does not know.}\end{defn}
As before, determinism is built into this definition insofar as $A$, $B$, and each member of $\mathcal{P}$ are required to be well-defined as functions, and locality is included in the requirement that $A$, $B$, and each member of $\mathcal{P}$ depend only on the measurement vector associated with the given measurement.  The idea behind this last requirement is that a local generalized hidden variable model must be one on which not only are Alice's measurement results independent of Bob's measurement settings, but also the projection scheme by which Alice translates her $X-$valued measurement results into integers $\pm1$ must be independent of Bob's measurement settings (and vice versa).

The relevant expectation values can now be seen to be the expectation values of the observables $A$ and $B$ \emph{after} they have been mapped to $\{-1,1\}$.  (We will use the symbol $\mathcal{E}_{g.h.v}$ to represent this new notion of expectation value.)  We can thus stipulate more clearly that, in order for a generalized hidden variable model $(\Lambda, X, A, B,\rho, \mathcal{P})$ to be satisfactory, it must be the case that for every choice of $\mathbf{a}$ and $\mathbf{b}$, the following hold:\footnote{I am now switching back to the notation used in section \ref{bell}, to emphasize the generality of these expressions. I am also explicitly indicating that the projection maps may depend on the measurement vectors.}
\begin{align}
\mathcal{E}_{g.h.v.}(A(\mathbf{a}))&=\int_{\Lambda}P_{\mathbf{a}}(A(\mathbf{a},\lambda))\rho(\lambda)d\lambda=0\\
\mathcal{E}_{g.h.v.}(B(\mathbf{b}))&=\int_{\Lambda}P_{\mathbf{b}}(B(\mathbf{b},\lambda))\rho(\lambda)d\lambda=0\\
\mathcal{E}_{g.h.v.}(A(\mathbf{a})B(\mathbf{b}))&=\int_{\Lambda}P_{\mathbf{a}}(A(\mathbf{a},\lambda))P_{\mathbf{b}}(B(\mathbf{b},\lambda))\rho(\lambda)d\lambda =-\mathbf{a}\cdot\mathbf{b}\label{biggie}
\end{align}
In these new terms, the model presented above is indeed a local, deterministic generalized hidden variable model.  But it does not reproduce the EPR-Bohm measurement results because it fails to satisfy Eq. \eqref{biggie}.

These considerations suggest a simple no-go result that helps focus attention on the case of a generalized hidden variable model.\footnote{I should emphasize that this result applies irrespective of $X$ and of $\mathcal{P}$, which means that it precludes models like the one Christian proposes in addition to the model I have discussed.  It also means that one cannot get around the central claim with a clever choice of $\mathcal{P}$.}  As noted above, the following is an immediate corollary of Theorem \ref{chsh}.
\begin{cor}\label{nogo2}\singlespacing
There does not exist a local, deterministic generalized hidden variable model of the EPR-Bohm experiment that reproduces the quantum mechanical expectation values.
\end{cor}
Proof. Let $(\Lambda, X, A, B,\rho, \mathcal{P})$ be a local, deterministic generalized hidden variable model.  Define two new observables $A',B':\mathbb{S}^2\times\Lambda\rightarrow\{-1,1\}$ as follows: for any $\mathbf{a},\mathbf{b}\in\mathbb{S}^2$ and any $\lambda\in\Lambda$, set $A'(\mathbf{a},\lambda)=P_{\mathbf{a}}(A(\mathbf{a},\lambda))$ and $B'(\mathbf{b},\lambda)=P_{\mathbf{b}}(B(\mathbf{b},\lambda))$.  Now $(\Lambda, A', B', \rho)$ is a hidden variable model (in the original sense of definition \ref{hvm}).  Thus, for any $\mathbf{a}$, $\mathbf{a}'$, $\mathbf{b}$, and $\mathbf{b}'$, the expectation values of $A'$ and $B'$ must satisfy the Clauser-Horne-Shimony-Holt inequality.  It follows, by direct substitution, that \begin{equation}
|\mathcal{E}_{g.h.v.}(A(\mathbf{a})B(\mathbf{b}))-\mathcal{E}_{g.h.v.}(A(\mathbf{a})B(\mathbf{b}'))+\mathcal{E}_{g.h.v.}(A(\mathbf{a}')B(\mathbf{b})) + \mathcal{E}_{g.h.v.}(A(\mathbf{a}')B(\mathbf{b}'))|\leq 2.
\end{equation}
As noted above, the quantum mechanical expectation values do not satisfy this inequality.  Thus no local, deterministic generalized hidden variable model can reproduce the quantum mechanical expectation values for an EPR-Bohm experiment.\hspace{.25in}$\square$

It follows that one cannot get around Bell-type theorems by considering more general representations of experimental outcomes.

\section{Final remarks}\label{conc}

The principal moral of this paper is that one cannot get around Bell-type arguments by generalizing the space of measurement outcomes, or equivalently, by representing the state of the particles after measurement with more sophisticated mathematical machinery.  The reason for this is that ultimately the measurement data that exhibits the Bell-type-theorem-violating correlations consists of sequences of elements of the set $\{-1,1\}$.  It is these sequences, or at least their expectation values, that a (generalized) hidden variable model of the EPR-Bohm experiment needs to explain in order to be satisfactory.  Corollary \ref{nogo2} shows how this is impossible, since the combination of any non-integer valued representation of measurement results with a map that identifies these with integers $\pm1$ simply recovers an observable in the standard sense of Bell-type theorems.  And we already know that a hidden variable model with standard, integer-valued observables cannot successfully reproduce the predictions of quantum mechanics.

This last point underscores the scope and generality of Bell-type theorems: they really do have the foundational importance typically attributed to them, contra the suggestion offered at the beginning of section \ref{uhoh} that Bell made a serious error in how he constrained the representation of EPR-Bohm measurement results.  The model and the no-go result I present here may help to clarify how it is that Bell-type theorems apply even to models that look quite different from the models that appear to be the target of standard Bell-type theorems.  But the present no-go result is at best a corollary of results that have been known for forty years.

Indeed, Bell understood very well that his argument was entirely independent of any concerns about what the mathematical and physical structure of the space of possible states $\Lambda$ might turn out to be---concerns of just the sort that motivate the model described in this paper and other similar models discussed recently.  As he put it,
\begin{quote}
...[Y]ou might still expect that it is a preoccupation with determinism that creates the problem.  Note well then that the following argument makes no mention whatever of determinism.  You might suspect that there is something specially peculiar about spin-1/2 particles.  ... So the following argument makes no reference to spin-1/2 particles, or any other particular particles.  Finally you might suspect that the very notion of particle ... has somehow led us astray.  Indeed did not Einstein think that fields rather than particles are at the bottom of everything?  So the following argument will not mention particles, nor indeed fields, nor any other particular picture of what goes on at the microscopic level.  Nor will it involve any use of the words ``quantum mechanical system", which can have an unfortunate effect on the discussion.  The difficulty is not created by any such picture or any such terminology.  It is created by the prediction about the correlations in the visible outputs of certain conceivable experimental set-ups. \citep[pg. 11]{Bell-BSNR}
\end{quote}
As Bell says, what matters is precisely the correlations that in fact obtain in certain experiments---correlations between sequences of results recorded in the standard way.  It is these correlations that we would want to explain with some successful model of the experiments.  And, as Bell showed, it is precisely these correlations that cannot be explained with any local, deterministic hidden variable model.

As a coda to the present discussion, it is interesting to note that if one began with a different version of Bell's theorem, in particular one of the probabilistic variety often associated with the Clauser-Horne inequality and with Jarrett's analysis of Bell-type results, the same basic issues arise.  In that setting one would have some joint distribution over what I call $A$ and $B$ above, which, in the ``generalized'' setting would be a probability measure on the space $X\times X$.  One could then claim to avoid Bell-type theorems by making the same sort of mistake made in Eq. \eqref{jcje}---namely taking the expectation value of the inner product (say) of the two elements of $X$ associated with each point of $X\times X$.  As with the model discussed here, one could certainly give the appearance of violating Bell-type theorems in this way.  But by the same considerations discussed in section \ref{nogo}, what one would really want would be the expectation value of the ordinary product after projection to $\{-1,1\}$.  And once again, we know this is impossible, precisely because of Bell-type theorems.

\section*{Acknowledgments}

I am grateful to Shelly Goldstein, David Malament, John Manchak, Florin Moldoveanu, Abner Shimony, Howard Stein, and an anonymous referee for helpful comments on a previous draft of this paper.  Thank you, too, to Jeff Barrett, Harvey Brown, David Malament, and Tim Maudlin for discussions on these and related topics and to David Hestenes for a helpful email exchange on Clifford algebras.  I am especially grateful to Joy Christian for an informative correspondence and for his kind help on this project despite our continued disagreement.  Finally, this paper has benefited from audiences at the Southern California Philosophy of Physics Group and the Maryland Foundations of Physics Group, and I am grateful to Christian W\"uthrich and Jeff Bub for inviting me to make these presentation.

\singlespacing

\end{document}